\documentstyle[12pt]{article}
\def\PRL #1 #2 #3{{\sl Phys. Rev. Lett.} {\bf#1} (#2) #3}
\def\NPB #1 #2 #3{{\sl Nucl. Phys.} {\bf B#1} (#2) #3}
\def\NPBFS #1 #2 #3 #4{{\sl Nucl. Phys.} {\bf B#2} [FS#1] (#3) #4}
\def\CMP #1 #2 #3{{\sl Commun. Math. Phys.} {\bf #1} (#2) #3}
\def\PRD #1 #2 #3{{\sl Phys. Rev.} {\bf D#1} (#2) #3}
\def\PLA #1 #2 #3{{\sl Phys. Lett.} {\bf #1A} (#2) #3}
\def\PLB #1 #2 #3{{\sl Phys. Lett.} {\bf #1B} (#2) #3}
\def\JMP #1 #2 #3{{\sl J. Math. Phys.} {\bf #1} (#2) #3}
\def\PTP #1 #2 #3{{\sl Prog. Theor. Phys.} {\bf #1} (#2) #3}
\def\SPTP #1 #2 #3{{\sl Suppl. Prog. Theor. Phys.} {\bf #1} (#2) #3}
\def\AoP #1 #2 #3{{\sl Ann. of Phys.} {\bf #1} (#2) #3}
\def\PNAS #1 #2 #3{{\sl Proc. Natl. Acad. Sci. USA} {\bf #1} (#2) #3}
\def\RMP #1 #2 #3{{\sl Rev. Mod. Phys.} {\bf #1} (#2) #3}
\def\PR #1 #2 #3{{\sl Phys. Reports} {\bf #1} (#2) #3}
\def\AoM #1 #2 #3{{\sl Ann. of Math.} {\bf #1} (#2) #3}
\def\UMN #1 #2 #3{{\sl Usp. Mat. Nauk} {\bf #1} (#2) #3}
\def\FAP #1 #2 #3{{\sl Funkt. Anal. Prilozheniya} {\bf #1} (#2) #3}
\def\FAaIA #1 #2 #3{{\sl Functional Analysis and Its Application} {\bf
#1} (#2) #3}
\def\BAMS #1 #2 #3{{\sl Bull. Am. Math. Soc.} {\bf #1} (#2)
#3} \def\TAMS #1 #2 #3{{\sl Trans. Am. Math. Soc.} {\bf #1} (#2) #3}
\def\InvM #1 #2 #3{{\sl Invent. Math.} {\bf #1} (#2) #3}
\def\LMP #1 #2 #3{{\sl Letters in Math. Phys.} {\bf #1} (#2) #3}
\def\IJMPA #1 #2 #3{{\sl Int. J. Mod. Phys.} {\bf A#1} (#2) #3}
\def\AdM #1 #2 #3{{\sl Advances in Math.} {\bf #1} (#2) #3}
\def\RMaP #1 #2 #3{{\sl Reports on Math. Phys.} {\bf #1} (#2) #3}
\def\IJM #1 #2 #3{{\sl Ill. J. Math.} {\bf #1} (#2) #3}
\def\APP #1 #2 #3{{\sl Acta Phys. Polon.} {\bf #1} (#2) #3}
\def\TMP #1 #2 #3{{\sl Theor. Mat. Phys.} {\bf #1} (#2) #3}
\def\JPA #1 #2 #3{{\sl J. Physics} {\bf A#1} (#2) #3}
\def\JSM #1 #2 #3{{\sl J. Soviet Math.} {\bf #1} (#2) #3}
\def\MPLA #1 #2 #3{{\sl Mod. Phys. Lett.} {\bf A#1} (#2) #3}
\def\JETP #1 #2 #3{{\sl Sov. Phys. JETP} {\bf #1} (#2) #3}
\def\JETPL #1 #2 #3{{\sl  Sov. Phys. JETP Lett.} {\bf #1} (#2) #3}
\def\PHSA #1 #2 #3{{\sl Physica} {\bf A#1} (#2) #3}
\def\CQG #1 #2 #3{{\sl Class. Quantum Grav.} {\bf #1} (#2) #3}

\begin{document}

\thispagestyle{empty}

\bigskip\bigskip\begin{center} {\Large
\bf {A generalized twistor dynamics of the D=3\\
SUSY systems.}}
\end{center}  \vskip 1.0truecm

\centerline{\bf Alexei Yu. Nurmagambetov and
Vladimir I. Tkach} \begin{center} {\it\small
National Science Center \\ "Kharkov Institute of Physics \& Technology"
\\310108, Kharkov} \end{center} \bigskip
\nopagebreak

{\small A generalization of the twistor shift procedure
to the case of superparticle interacting with the background D=3 N=1
Maxwell and D=3 N=1 supergravity supermultiplet is considered. We
investigate twistor shift effects and discuss the structure of the
resulting constraint algebra.}

\vspace{1cm}

\section{Introduction}

The hope to solve the problem of the covariant quantization of
Green -- Schwarz superstrings is connected now with a success and further
development of the twistor -- like and relative
approaches \cite{penr} to the theory of extended supersymmetric objects.
In this respect of great importance is the studying of
twistor~--~like formulations of superparticles which are a limiting case
of superstrings theory.

Two attractive superparticle formulations in the framework of the
twistor~--~like approach intensively investigated nowadays are the
Ferber~--~Shirafuji \cite{fer,schir} and the
Sorokin~--~Tkach~--~Volkov~--~Zheltukhin (STVZ) \cite{stv} formulation.

The STVZ action is a version of a massless relativistic superparticle with
intristic incorporation  of twistor -- like commuting spinors into the
theory.

The presence of the twistor -- like spinors gave the possibility of
decomposing  the set of constraints of the theory onto the first and
second class without introducting the additional "Stuckelberg" variables
\cite{twes,nis} by  projecting it onto the twistor directions
\cite{stv}  and solved the problem of the infinite reducibility of
the $\kappa$ -- symmetry by replacing the former with the
superconformal worldline supersymmetry \cite{vz} being irreducible by
definition.

Several versions of twistor -- like supersymmetric particles and heterotic
strings based on the STVZ action have been constructed in D=3,4 and 6
space~--~time dimensions \cite{stv,pash,ds,gal,ivanov}.
However, the generalization of the STVZ action to the case of a D=10
superparticle posseses another infinite reducible symmetry \cite{gal}
that can be an obstacle  for applying the BRST~--~BFV~--~BV
quantization scheme.

Perhaps, the solution to this problem can be found by considering
the Ferber \cite{fer} and Shirafuji \cite{schir} twistor superparticle
action classically equivalent to the STVZ action.
Unfortunately, a worldline supersymmetric version of the
Ferber~--~Shirafuji action \cite{vz} is part of a more general action
describing the so called spinning superparticle \cite{spinsup} and does
not describe the usual N=1 Brink~--~Schwarz superparticle, but a model
with not well defined physical content because the target space is not the
conventional superspace, but one with additional $\theta$~--~translation.
However, we may consider a sector of this theory corresponding to the N=1
Brink -- Schwarz superparticle \cite{bs} and containing a particular
solution to the equations of motion analogous to that considered in
\cite{vz}. The problem of the appropriate worldline supersymmetrization
of the Ferber~--~Schirafuji action has been solved recently in
\cite{bnsv}.

In both, the Ferber -- Shirafuji and STVZ action, twistor -- like
variables are non-propagating auxiliary degrees of freedom. But we can
"animate" their by adding new terms depending on twistor -- like
variables and their proper~--~time derivatives. The terms of such kind can
arise as a result of particle interaction with quantum
fields. Since twistor -- like variables are commuting spinors it seems
that the introduction of new terms breaks the Pauli spin~--~statistics
theorem.  But in the paper  \cite{sorv}, where group
-- theoretical description of semion dynamics in D=3 was considered, it
was shown that an "animation" of twistor~--~like variables in the case of
the STVZ action describing a massive particle leads to arising
states with spin 1/4 and 3/4. Thus, the introduction of the new term
can drastically change the situation and lead to models for
describing more general quantum objects.

On the other hand the twistor dynamics described by the
Ferber~--~Shirafuji action can be modified by
analogous terms \cite{sstv}. In the case of a free particle
due to a so called twistor shift procedure a generalized dynamics
is equivalent to the standard one, but in the case of particle interaction
with background fields it leads to a modification of the interaction
which becomes nonminimal and is characterized by infinite series of terms.
The modified interaction contains a field
strength tensor and its higher derivatives and becomes nonlocal. In the
case of a D=3 Maxwell field background the first term in this series
describes a particle possessing an anomalous magnetic moment. As result,
one arrives at a model relevant to  Chern~--~Simons
systems and self -- interacting anyons (see,for example, references
in \cite{sorv}).

Such fundamental notions as locality and causality are
basically connected with the concept of background fields. There are
two principal kinds of locality and causality, namely the worldline
(worldsheet) and target space ones. The string ideology is based on the
worldsheet locality while field theory deals with locality in
target space. Since all physical fields in the framework of string concept
are nothing but superstring oscillations it is very important to
understand the connection between worldsheet and target space locality.
String interaction generate an effective interaction of fields which is
characterized by terms with higher derivatives of
the field potentials, thus leading to non-local interactions.
The presence of these new terms affects the structure of
field equations of motion, integrability conditions etc. An anologous
situation is reproduced in the generalized twistor dynamics.

{}From this point of view the main motivation of our research is to
study the twistor shift procedure in the case of a superparticle
interacting with background superfields, its effects and structure of the
constraint algebra.

The article is organized as follows. Section 2 briefly describes
twistor dynamics and its generalization.  In section 3 we investigate
twistor shift effects in the system describing superparticle interacting
with D=3 background Maxwell supermultiplet and discuss the
constraint algebra. In section 4 we consider as more complicated case
superparticle in D=3 supergravity background.The discussion of the
obtained results is given in the section Conclusion.

\section{Twistor dynamics and its generalization.}

Twistor -- like formulation of D=3 particle is
based on the following action for D=3 massless spinless relativistic
particle proposed by Sorokin, Tkach, Volkov and Zheltukhin \cite{stv}

\begin{equation}
S_{S.T.V.Z}=\int\,dt\,p_{m}(\dot{x}^{m}-\lambda\gamma^{m}\lambda)
\end{equation}
This action reproduces the Cartan -- Penrose momentum representation
\cite{penr}

\begin{equation}\label{3}
p_{m}=\lambda\gamma_{m}\lambda
\end{equation}
as constraint solution
\begin{equation}\label{2.2}
       p^2=0
\end{equation}
on the equation of motion.

With the help of the representation (2) it is easy to prove classical
equivalence of the action (1) to the twistor particle action proposed by
Ferber and Shirafuji \cite{fer,schir}:
\begin{equation}\label{8}
S_{twistor}=\int\,dt\,\lambda\gamma_{m}\lambda\dot{x}^{m}
\end{equation}
where as well as in the action (1) $\lambda$ is the commuting Majorana
spinors.

The further generalization of the massless relativistic particle twistor
dynamics is connected with ref. \cite{sstv} where with the aim of
"animation" of the twistor variables being non-propagating auxiliary
degrees of freedom the action (4) was modified by addition of a term
depending on the twistor variables and their proper -- time derivatives
the simplest variant of which has the following form:
\begin{equation}\label{10}
S_{additional}=\int\,d\tau\,l\lambda_{\alpha}\dot{\lambda}_{\beta}
\varepsilon^{\alpha\beta}
\end{equation}

The Hamilton analysis of the system described by the action
\begin{equation}
S_{general}=S_{twistor} + S_{additional}
\end{equation}
shows that due to the existence of the nontrivial transformation to the
new space -- time coordinates,
\begin{equation}
  \hat{x}^{\alpha\beta}=x^{\alpha\beta}+{l\over 2(\lambda\mu)}
  (\lambda^{\alpha}\mu^{\beta}+\lambda^{\beta}\mu^{\alpha})
\end{equation}
$$x^{\alpha\beta}\equiv x^{m}{(\gamma_{m})}^{\alpha\beta}$$
containing parameter $l$ with the dimension of length, a generalized
twistor dynamics in the free case is equivalent to
the standard one, i.e.
\begin{equation}
S_{general}=\int\,d\tau\,\lambda_{\alpha}\lambda_{\beta}
\dot{\hat{x}}^{\alpha\beta}
\end{equation}
The transformation (7) is usually called the twistor shift transformation.

The situation drastically changes in the case of particle interaction with
an external background field since the field potentials depending on the
space -- time coordinates at the transition to the new space -- time
coordinates $\hat{x}$ reduce to the arising of the infinite power series
of the nonlocality parameter with dimension of length, containing
field strength tensor and its higher derivatives. Thus, in the framework
of the generalized dynamics the interaction changes from minimal
scheme to nonminimal.

In the case of particle interaction with the
Maxwell background field the interaction modification in the first power
of $l$ has the following form
\begin{equation}
{\cal{A}}_{m}\rightarrow{\cal{A}}_{m} + l\varepsilon_{mnk}F^{nk}
\end{equation}
where $F^{nk}$ is the Maxwell strength tensor, $\varepsilon_{mnk}$ is the
Levi -- Chivita tensor, and is related to the interaction of
particle having an anomalous magnetic moment with Maxwell field. The
theories of such kind are intensively investigated nowadays in connection
with possible applications to the theory of hight temperature
superconductivity and fractional quantum Hall effect.


\section{Supertwistor shift in the presence of the background Maxwell
field.}

A generalized dynamics of superparticle interacting with the Maxwell
background superfield is described by the following action
\begin{equation}
S=\int\,d\tau\,d\eta\,(iE^{-1}DE^{\alpha}DE^{\beta}(\gamma_{m})_{\alpha\beta}
DE^{m}+lE^{-1}\dot{E^{\alpha}}DE_{\alpha}+DE^{A}\cal{A_{A}})
\end{equation}
where
$D=\partial_{\eta}+i\eta\partial_{\tau}$
is the covariant derivative of the "small" world -- line SUSY parametrized
by proper -- time $\tau$ and its grassman superpartner $\eta$
$$DE^{A}=(Dz^{M})E_{M}^{A}$$
$E_{M}^{A}(z)$ is the standard superdreinbein of the flat superspace with
coordinates $z^{M}~=~(X^{m},\Theta^{\alpha})$
\begin{equation} X^{m}=x^{m}+i\eta\chi^{m};\\
\Theta^{\alpha}=\theta^{\alpha}+ \eta\lambda^{\alpha}
\end{equation}
being a scalar under "small" SUSY transformations.

$E^{-1}=e^{-1}-i\eta{\hat{\psi}}/{e^{2}}$
is the analog of world -- line supergravity, $\cal{A}$ is the background
Maxwell superfield and $l$ is the nonlocality parameter having the
dimension of length. Such form of the action is very convenient for the
application to the case of background supergravity considered in the next
section.

The action (10) is closely related to one proposed in ref. \cite{ds}. In
the paper of Delduc and Sokatchev the conventional background constaints
( namely (15), (16)) playing the role of the integrability conditions
and maintaining original theory symmetries in the presence of background
interaction was obtained in elegant manner. The presence of the additional
term does not change this situation that gives a possibility to consider
usual constraints imposed to the Maxwell and supergravity background
superfield.

Twistor shift is generated by transition to the new space -- time variables
\begin{equation}
  \hat{X}^{\alpha\beta}=X^{\alpha\beta}+{l\over2{\tilde{\mu}}D\Theta}
  (\tilde{\mu}^{\alpha}D\Theta^{\beta}+\tilde{\mu}^{\beta}D\Theta^{\alpha})
\end{equation}
where
$\tilde{\mu}^{\alpha}=\mu^{\alpha}+{\eta}d^{\alpha}$
is the even superfield containing the second twistor half $\mu$ and its
superpartner $d$, moreover,
$$ \tilde{\mu}_{\alpha}=iX_{\alpha\beta}D\Theta^{\beta}$$

By using the equation of motion in the first order over $l$ after
the twistor shift we obtain the following action modulo term
$D\Theta^{\alpha}{\cal{A}}_{\alpha}$ which we shall consider below
\begin{equation}
\tilde{S}=\int\,d\tau\,d\eta\,
(iE^{-1}D\Theta\gamma_{m}{D\Theta}D\hat{X}^{m}+\hat{\Omega}^{m}{\cal A}_{m}
+{1\over2}
l\hat{\Omega}^{\alpha\beta}{(\sigma^{mn})}_{\alpha\beta}{\cal F}_{mn})
\end{equation}
where $\hat{\Omega}^{m}$ is the Cartan form invariant under local "small"
and global "big" SUSY transformations
$$\hat{\Omega}^{m}=D\hat{X}^{m}+i\Theta\gamma^{m}D\Theta+
iD\Theta\gamma^{m}\Theta$$.

In D=3 the vector field is a part of spinor supermultiplet
\begin{equation}
  {\cal
  A}_{\alpha}=r_{\alpha}+B\Theta_{\alpha}+V_{\alpha\beta}\Theta^{\beta}+
  h_{\alpha}\Theta\Theta
\end{equation}
where
$V_{\alpha\beta}=V^{m}{(\gamma_{m})}_{\alpha\beta}$
and $h$ is the vector gauge field and its superpartner, respectively.

The Wess -- Zumino gauge $r=B=0$ and conventional constraints
\begin{equation}
  {\cal F}_{\alpha\beta}(\hat{X},\Theta)=0
\end{equation}
\begin{equation}
  T^{a}_{\alpha\beta}(\hat{X},\Theta)=2i{\gamma^{a}}_{\alpha\beta}
\end{equation}
extract irreducible submultiplet of physical fields, after that the action
(13) may be present in the form of
$$
\tilde{S}=\int\,d\tau\,d\eta\,iE^{-1}D\Theta\gamma_{m}{D\Theta}D\hat{X}^{m}
-{i\over2}\hat{\Omega}^{m}(V_{m}+{1\over2}\Theta_{\beta}(\gamma_{m})^
{\beta\alpha}h_{\alpha}-{i\over2}\Theta\Theta{\varepsilon^{kl}}_{m}F_{kl})$$
\begin{equation}
+i{1\over2}l
\varepsilon^{nmk}\hat{\Omega}_{k}(F_{nm}+\Theta_{\alpha}(\gamma_{n})^
{\alpha\beta}\partial_{m}h_{\beta}-{i\over2}\Theta\Theta{\varepsilon^{pl}}_{m}
\partial_{n}F_{pl})
\end{equation}
where $\varepsilon_{mnk}$ is the Levi -- Chivita tensor;$F^{mn}$ is the
Maxwell field strength tensor.

Integrating over $\eta$ and fixing solution of the equation of motion
$${\delta{L}\over{\delta{\hat{\psi}}}}=0$$
in the form of
   \begin{equation}
   \chi^{m}=-(\theta\gamma^{m}\lambda+\lambda\gamma^{m}\theta)
   \end{equation}
we come to the component form of the Lagrangian (17). Such choosing
of the solution connected with the first term of the action (17) has been
considered by Volkov and Zheltukhin \cite{vz}.

It is easy to see writing down the remaining term of the action (10) into
the component form and using the equation of motion
$${\delta{L}\over{\delta{e}}}=0$$
that this one duplicates some terms of the action (17) so that full
Lagrangian after the twistor shift appears as
$$
\tilde{L}=-(e^{-1}\lambda\gamma_{m}\lambda-(V_{m}+{1\over2}\theta_{\beta}
\gamma_{m}^{\beta\alpha}h_{\alpha}-{i\over4}\theta\theta{\varepsilon^{kl}}_{m}
F_{kl})$$
\begin{equation}
+l{\varepsilon^{nk}}_{m}(F_{nk}+{1\over2}\theta_{\alpha}\gamma_{n}^
{\alpha\beta}\partial_{k}h_{\beta}-{i\over4}\theta\theta{\varepsilon^{pl}}_
{k}\partial_{n}F_{pl}))
{\dot{\omega}}^{m}
\end{equation}
where $\omega^{m}$ is the Cartan differential form
\begin{equation}
d\omega^{m}=d\hat{x}^{m}-i\theta\gamma^{m}d\theta+id\theta\gamma^{m}\theta
+2\lambda\gamma^{m}{\lambda}d\tau
\end{equation}

Lagrangian (19) is invariant under the following transformations of the
global SUSY on the mass -- shell
$$
  {\delta}V_{m}=-\varepsilon_{\beta}(\gamma_{m})^{\beta\alpha}h_{\alpha};
 \  {\delta}\Theta_{\beta}=\varepsilon_{\beta};
  \
{\delta}h_{\alpha}={i\over2}\varepsilon^{\beta}(\sigma^{nm})_{\beta\alpha}
F_{nm}
$$
with odd parameter $\varepsilon$.

Now let us consider the algebra of the constraints obtaining
from the action (19). In the case of usual Brink~--~Schwarz
superparticle this one has the following form:
$$\{d_{\alpha}, d_{\beta}\}=-2{(\gamma^{a})}_{\alpha\beta}p_{a}$$
where $d_{\alpha}$ and $d_{\beta}$ are the second class constraints and
$p_{a}$ is the covariant superparticle momentum transforming under the
quantization procedure into the covariant vector derivative. In our case
$$ d_{\alpha}=P_{{\theta}^{\alpha}}-(e^{-1}\lambda\gamma_{m}\lambda-
{\tilde{V}}_{m})\gamma^{m}\theta_{\alpha}$$
where
$$\tilde{V}_{m}=V_{m}+{1\over2}\theta_{\beta}
\gamma_{m}^{\beta\alpha}h_{\alpha}-{i\over4}\theta\theta{\varepsilon^{kl}}_{m}
F_{kl}$$
$$+l{\varepsilon^{nk}}_{m}(F_{nk}+{1\over2}\theta_{\alpha}\gamma_{n}^
{\alpha\beta}\partial_{k}h_{\beta}-{i\over4}\theta\theta{\varepsilon^{pl}}_
{k}\partial_{n}F_{pl})$$
and it is easy
to see that the algebra of constraints does not change upon applying the
twistor shift procedure in the first order in $l$ expansion with vector
covariant derivative having modification like (9), moreover, we hope that
it is remain true for any order of $l^{n}$ expansion.  The main reason of
preserving the algebraic structure is the conventional constraint (15)
reproducing the integrability condition.

Thus, the twistor shift procedure allows to get rid of undesirable term
$\dot{\theta}\dot{\theta}$ breaking the spin -- statistic connection under
turning into the field theory. This leads to arising in bosonic sector an
infinite series in the parameter $l$ containing Maxwell
strength tensor and its higher derivatives. The fermionic sector present
in the case of superparticle is also modified by terms containing
derivatives from a gauge field superpartner. The algebra of second
class constraints does not change upon the twistor shift procedure due to
imposing conventional constraint playing the role of integrability
condition.


\section{Supertwistor shift in the presence of the background
supergravity.}

The more complicated case is, of course, the case of the supergravity
background.  We would like firstly to discuss this one in general and then
to apply this consideration to our approach.

The spinorial covariant derivatives of the D=3 N=1 supergravity in the
Wess~--~Zumino gauge have the following form \cite{bg,gat,wz}:
\begin{equation}
D_{\alpha}=\partial_{\alpha}-i{1\over2}{(\gamma^{0}\gamma^{a}\Theta)}_{\alpha}
\nabla_{a}+{1\over4}\Theta\Theta{(\gamma^{0}\hat{\nabla})}_{\alpha}
\end{equation}
where
$$\nabla_{a}=\psi^{\mu}_{a}\partial_{\mu}+e^{m}_{a}\partial_{m}+
{1\over2}\omega^{kp}_{a}M_{kp}$$
\begin{equation}
{\hat{\nabla}}^{\alpha}=\nabla^{\alpha}+{1\over4}\omega^{kp}_{a}
{\varepsilon_{kp}}^{d}{(\gamma^{a}\gamma_{d}\partial)}^{\alpha}
+{1\over2}{(\gamma^{a}\gamma^{b}\psi_{a})}^{\alpha}\nabla_{b}
\end{equation}
$$\nabla^{\alpha}={(\phi\gamma^{0})}^{\alpha\mu}\partial_{\mu}+
\chi^{{\alpha}a}\nabla_{a}+{1\over2}\zeta^{{\alpha}kp}M_{kp}$$
In our notations $\partial_{m}={\partial}/{\partial}X^{m}$,
$\partial_{\alpha}={\partial}/{\partial\Theta^{\alpha}}$,
$(\gamma^{0})_{\alpha\beta}=\varepsilon_{\alpha\beta}$ in Majorana
representation of the $\gamma$ -- matrices (see Appendix), ${\varepsilon_
{kp}}^{d}$ is the Levi -- Chivita tensor;
$e^{m}_{a}$,
$\omega_{a}^{kp}$, $\psi_{a}^{\mu}$ are the components of the dreinbein,
connection and gravitino respectively; the structure group of the tangent
space is the $SL(2,R)$ with the Lorentz generators $M_{kp}$. The remaining
fields provide closing the SUSY algebra on the "off -- shell".

With the help of the expression (21) and (22) we can restore the
superdreinbein components by using of
\begin{equation}
D_{A}=E_{A}^{M}\partial_{M}+{W_{A}}^{kp}M_{kp}
\end{equation}
where indices $A=\{a,\alpha\}$ take a value of the vector and spinor
indices in the tangent "flat" superspace, $M=\{\underline{m},\mu\}$ are
analogous indices of the "curve" superspace, ${W_{A}}^{kp}$ is the
superconnection taking value in the structure group.

As in the case of the Maxwell background it is necessary to impose the
conventional constraints on the curvature and torsion consistent with the
Wess~--~Zumino gauge \cite{wz}:
$${\cal{R}}_{\alpha\beta}(X,\Theta)=0$$
$$T_{\alpha\beta}^{a}(X,\Theta)=-{i\over2}{\gamma^{a}}_{\alpha\beta}$$
so that
\begin{equation}
\{D_{\alpha},D_{\beta}\}=-i\gamma^{a}_{\alpha\beta}D_{a}
\end{equation}

Then from (21) and (22) we obtain
$$E_{\alpha}^{\mu}=\delta_{\alpha}^{\mu}-i{1\over2}(\gamma^{0}\gamma^{a}
\Theta)_{\alpha}\psi_{a}^{\mu}+{1\over4}\Theta\Theta(\gamma^{0})_{\alpha\beta}
\{(\phi\gamma^{0})^{\beta\mu}$$
\begin{equation}
+{1\over2}(\gamma^{a}\gamma^{b}\psi_{a})^{\beta}
\psi_{b}^{\mu}+{1\over4}\omega_{a}^{kp}{\varepsilon_{kp}}^{d}(\gamma^{a}
\gamma_{d})^{\beta\mu}\}
\end{equation}
$$E_{\alpha}^{\underline{m}}=-i{1\over2}(\gamma^{0}\gamma^{a}\Theta)_{\alpha}
e_{a}^{\underline{m}}+{1\over8}\Theta\Theta(\gamma^{0})_{\alpha\beta}
(\gamma^{a}\gamma^{b}\psi_{a})^{\beta}e_{b}^{\underline{m}}$$
and from (24) and
$$
D_{a}={E_{a}}^{\underline{m}}\partial_{\underline{m}}+{E_{a}}^{\mu}
\partial_{\mu}+{W_{a}}^{kp}M_{kp}
$$
we get the expression for the remaining superdreinbein components:
$$E_{a}^{\underline{m}}=e_{a}^{\underline{m}}+{1\over4}i\Theta\gamma_{a}
\psi^{\underline{m}}
-{1\over8}\Theta\Theta(\gamma^{b})_{\alpha\gamma}\psi_{b}^{\gamma}(\gamma^{c})
^{\alpha\varepsilon}\psi_{a\varepsilon}e_{c}^{\underline{m}}$$
$$+{1\over4}\Theta\Theta{\phi}e_{a}^{\underline{m}}-{1\over8}\Theta\Theta
\omega_{a}^{kp}Tr(\sigma_{kp}\gamma^{c})e_{c}^{\underline {m}}$$
$$E_{\underline{n}}^{\mu}=\psi_{\underline{n}}^{\mu}+{1\over4}i\Theta_{\alpha}
(\gamma^{a})^{\alpha\beta}\psi_{\underline{n}\beta}\psi_{a}^{\mu}-{1\over4}
i\Theta_{\alpha}(\sigma_{kp})^{\alpha\mu}\omega_{\underline{n}}^{kp}$$
\begin{equation}
+{1\over8}\Theta\Theta(\gamma^{a})_{\alpha\gamma}\psi_{a}^{\gamma}\omega_
{\underline{n}}^{kp}(\sigma_{kp})^{\alpha\mu}+{1\over4}\Theta\Theta\phi
\psi^{\mu}_{\underline{n}}-{1\over8}\Theta\Theta\omega_{\underline{n}}^{kp}
Tr(\sigma_{kp}\gamma^{a})\psi_{a}^{\mu}
\end{equation}
The last terms of these exrpessions vanish by virtue of
$Tr(\sigma_{kp}\gamma^{c})=0$, where $Tr$ denotes the trace and
$\sigma_{kp}$ is the antisymmetric product of Dirac matrices (see
Appendix).

Since the twistor shift procedure is carried out on the "mass -- shell"
we can neglect the auxiliary scalar field contribution.

The action describing superparticle in the gravitational background field
has the following form \cite{ds}:
\begin{equation}
S=-i\int\,d\tau\,d\eta\,E^{-1}DE^{\alpha}DE^{\beta}(\gamma_{m})_{\alpha\beta}
DE^{m}
\end{equation}
Using $DE^{A}=(Dz^{M})E_M^A$ we obtain the following expression for the
Lagrangian:
$$L=D\Theta^{\lambda}E_{\lambda}^{\alpha}D\Theta^{\gamma}E_{\gamma}^{\beta}
DX^{\underline{k}}E_{\underline{k}}^{m}(\gamma_{m})_{\alpha\beta}+2D\Theta^
{\lambda}E_{\lambda}^{\alpha}DX^{\underline{n}}E_{\underline{n}}^{\beta}D
\Theta^{\gamma}E_{\gamma}^{m}(\gamma_{m})_{\alpha\beta}$$
\begin{equation}
+(\gamma_{m})_{\alpha\beta}(2D\Theta^{\lambda}E_{\lambda}^{\alpha}DX^
{\underline{n}}E_{\underline{n}}^{\beta}DX^{\underline{k}}E_{\underline{k}}^
{m}+DX^{\underline{m}}E_{\underline{m}}^
{\alpha}DX^{\underline{n}}E_{\underline{n}}^{\beta}D\Theta^{\lambda}E_
{\lambda}^{m})
\end{equation}
(By using the superreparametrization we always can choose $E^{-1}=1$).

It is easy to prove that the last two terms of the (28) vanish (for this
aim it is nessesary to write down these terms in the component form) and
after that, using (25) and (26) we can find the remaining Lagrangian
$$L=D\Theta^{\lambda}D\Theta^{\gamma}DX^{\underline{n}}(\gamma_{m})_
{\alpha\beta}\{\delta_{\lambda}^{\alpha}\delta_{\gamma}^{\beta}e^{m}_
{\underline{n}}+{1\over4}i\delta_{\lambda}^{\alpha}\delta_{\gamma}^{\beta}
\Theta\gamma^{m}\psi_{\underline{n}}-i(\gamma^{k}\Theta)_{\gamma}\psi_{k}^
{\beta}\delta_{\lambda}^{\alpha}e_{\underline{n}}^{m}$$
$$+{1\over2}\Theta\Theta\delta_{\lambda}^{\alpha}({1\over2}(\gamma^{a}
\gamma^{b}\psi_{a})_{\gamma}\psi_{b}^{\beta}e_{\underline{n}}^{m})-{1\over8}
\delta_{\lambda}^{\alpha}\delta_{\gamma}^{\beta}\Theta\Theta(\gamma^{b})
_{\alpha\gamma}\psi_{b}^{\gamma}(\gamma_{c})^{\alpha\varepsilon}\psi^{m}
_{\varepsilon}e_{\underline{n}}^{c}\}$$
\begin{equation}
+2D\Theta^{\lambda}D\Theta^{\gamma}DX^{\underline{n}}(\gamma_{m})_{\alpha
\beta}\{-{i\over2}\delta_{\lambda}^{\alpha}\psi_{\underline{n}}^{\beta}
(\gamma^{m}\Theta)_{\gamma}+{1\over8}\delta_{\lambda}^{\alpha}(\Theta\gamma^
{c}\psi_{\underline{n}})\psi_{c}^{\beta}(\gamma^{m}\Theta)_{\gamma}
\end{equation}
$$-{1\over4}(\gamma^{k}\Theta)_{\lambda}\psi_{k}^{\alpha}
\psi_{\underline{n}}^{\beta}(\gamma^{m}\Theta)_{\gamma}
+{1\over8}\Theta\Theta\delta_{\lambda}^{\alpha}\psi_{\underline{n}}^{\beta}
(\gamma^{a}\gamma^{m}\psi_{a})_{\gamma}\}$$
With help of the evident identity
$$\Theta^{\alpha}\Theta^{\beta}=-{1\over2}\varepsilon^{\alpha\beta}\Theta
\Theta$$
the Lagrangian (29) is reduced to the following form
\begin{equation}
L=D\Theta^{\lambda}D\Theta^{\gamma}DX^{\underline{n}}(\gamma_{m})_{\alpha
\beta}\{\delta_{\lambda}^{\alpha}\delta_{\gamma}^{\beta}e_{\underline{n}}^{m}
+{1\over4}i\delta_{\lambda}^{\alpha}\delta_{\gamma}^{\beta}\Theta\gamma^{m}
\psi_{\underline{n}}
\end{equation}
$$+{1\over4}\Theta\Theta\delta_{\lambda}^{\alpha}(\gamma^{a}
\gamma^{m}\psi_{a})_{\gamma}\psi_{\underline{n}}^{\beta}\}$$
and describes the superparticle interacting with the
background D=3 N=1 supergravity.

The generalization of the dynamics
is achieved by introduction of the additional term
\begin{equation}
L_{additional}=-il{\dot{E}}^{\alpha}DE_{\alpha}
\end{equation}
having an extended expression in the form of
\begin{equation}
L_{additional}=-il\{({\dot{z}}^{N}Dz_{M}(-)^{(\alpha+N)(M+1)}E_{N}^{\alpha}
E_{\alpha}^{M}
\end{equation}
$$+(-)^{\alpha(N+1)}(Dz^{N})(Dz_{M})E_{\alpha}^{M}(DE_{N}^
{\alpha})\}$$

Since the expression (32) contains $z^{N}$ we can consider the
contribution of $X^{m}$ variables, but for investigation of our
system it is neseccary to choose namely $z^{\mu}\equiv\Theta^{\mu}$
(accounting $X^{m}$ leads to considering the spinning superparticle
\cite{spinsup} interacting with the background supergravity).
Then
\begin{equation}
L_{additional}=-il({\dot{\Theta}}^{\nu}D\Theta_{\mu}E_{\nu}^{\alpha}E_{\alpha}
^{\mu}+D\Theta^{\nu}D\Theta_{\mu}DX^{\underline{k}}({\tilde{\Omega}}_
{\underline{k}})_{\nu}^{\mu})
\end{equation}
where $({\tilde{\Omega}}_{\underline{k}})_{\nu}^{\mu}$ is the object of
unholonomity having only the following non-vanishing components
\begin{equation}
({\tilde{\Omega}}_{\underline{k}})_{\nu}^{\mu}=({\tilde{\Omega}}^{a}_{bc})
e_{a\underline{k}}(-{1\over2})(\sigma^{bc})_{\nu}^{\mu}
\end{equation}

Imposing the additional torsion constraint
\begin{equation}
T^{a}_{bc}=0
\end{equation}
makes it possible to express the superconnection $\Omega$ through
the $\tilde{\Omega}$ as:

\begin{equation}
\tilde{\Omega}^{a}_{bc}=-2{\Omega^{a}}_{[bc]}
\end{equation}
where the braces $[\dots]$ denote an antisymmetrization.

The connection of the unholonomity object with the
superdreinbein components is well known:
\begin{equation}
\tilde{\Omega}^{A}_{BC}=(-)^{B(M+C)}V_{C}^{M}V_{B}^{N}(E^{A}_{M,N}-(-)^{MN}
E^{A}_{N,M})
\end{equation}
where $E^{A}_{M,N}={\partial{E^{A}_{M}}}/{\partial{z^{N}}}$.

Then
$$
\tilde{\Omega}^{a}_{bc}=~e_{c}^{\underline{m}}e_{b}^{\underline{n}}
(\partial_{\underline{n}}e^{a}_{\underline{m}}+{1\over4}i\theta\gamma^{a}
\partial_{\underline{n}}\psi_{\underline{m}}-{\scriptstyle(\underline{m}
\longleftrightarrow\underline{n})})$$
$$
+{1\over4}i\theta\gamma_{b}\psi^{\underline{n}}e_{c}^{\underline{m}}
(\partial_{\underline{n}}e^{a}_{\underline{m}}-{\scriptstyle(\underline{m}
\longleftrightarrow\underline{n})})
+{1\over4}i\theta\gamma_{b}\psi^{\underline{n}}e_{c}^{\underline{m}}
({1\over4}i\theta\gamma_{a})(\partial_{\underline{n}}\psi_{\underline{m}}-
{\scriptstyle(\underline{m}\longleftrightarrow\underline{n})})$$
$$
+{1\over4}i\theta\gamma_{c}\psi^{\underline{m}}e_{b}^{\underline{n}}
(\partial_{\underline{n}}e^{a}_{\underline{m}}-{\scriptstyle(\underline{m}
\longleftrightarrow\underline{n}))})
+{1\over4}i\theta\gamma_{c}\psi^{\underline{m}}e_{b}^{\underline{n}}({1\over4}
i\theta\gamma_{a})(\partial_{\underline{n}}\psi_{\underline{m}}-
{\scriptstyle(\underline{m}\longleftrightarrow\underline{n})})$$
$$
-{1\over16}(\theta\gamma_{c}\psi^{\underline{m}})(\theta\gamma^{b}
\psi^{\underline{n}})
(\partial_{\underline{n}}e^{a}_{\underline{m}}-{\scriptstyle(\underline{m}
\longleftrightarrow\underline{n})})+{1\over4}\theta\theta
e_{c}^{\underline{m}}e_{b}^{\underline{n}}({1\over2}(\gamma^{b}
\gamma^{a}\psi_{b})_{\beta}(\partial_{\underline{n}}\psi^{\beta}_
{\underline{m}}-{\scriptstyle(\underline{m}\longleftrightarrow
\underline{n})}))$$
\begin{equation}
-{1\over8}\theta\theta\{e_{c}^{\underline{m}}(\gamma^{a}
\gamma_{b}\psi_{a})^{\beta}\psi_{\beta}^{\underline{n}}+e_{b}^{\underline{n}}
(\gamma^{a}\gamma_{c}\psi_{a})^{\beta}\psi_{\beta}^{\underline{m}}\}
(\partial_{\underline{n}}e^{a}_{\underline{m}}-{\scriptstyle(\underline{m}
\longleftrightarrow\underline{n})})
\end{equation}

Now let us turn back to the (30), (33) and carry out the twistor shift by
replacing of the space -- time coordinates \begin{equation}
\hat{X}_{\underline{m}}=X_{\underline{m}}+{l\over2D\Theta\tilde{\mu}}D\Theta^
{\alpha}{(\gamma_{a})_{\alpha}}^{\beta}\tilde{\mu}_{\beta}E^{a}_{\underline
{m}}
\end{equation}
that leads to arising a new term
$$ -i{1\over2}l{\varepsilon_{\underline{k}}}^{\underline{nl}}\partial_
{\underline{n}}E_{\underline{l}}^{m}(\gamma_{m})_{\alpha\beta}D\Theta^
{\alpha}D\Theta^{\beta}DX^{\underline{k}}$$

Then after cumbersome calculations using Volkov -- Zheltukhin
solution \cite{vz} we obtain the Lagrangian describing a generalized
twistor dynamics of superparticle interacting with the D=3 N=1
supergravity background field
$$L^{shift}_{SG}=\lambda^{\alpha}\lambda^{\beta}(\gamma_{a})_{\alpha\beta}
\{e^{a}_{\underline{m}}+{1\over4}i\theta\gamma^{a}\psi_{\underline{m}}
+{1\over4}\theta\theta(\gamma^{b}\gamma^{a}\psi_{b})_{\gamma}\psi^{\gamma}_
{\underline{m}}$$
\begin{equation}\qquad
+{1\over4}l{\varepsilon_{\underline{m}}}^{\underline{n}\underline{l}}
(T^{a}_{\underline{n}\underline{l}}+{1\over4}i(\theta\gamma^{a})_{\rho}
T^{\rho}_{\underline{n}\underline{l}}+{1\over4}\theta\theta(\gamma^{b}
\gamma^{a}\psi_{b})_{\rho}T^{\rho}_{\underline{n}\underline{l}})
\}{\dot{\omega}}^{\underline{m}}
\end{equation}
Here all fields depend on the new space -- time variables $\hat{x}$,
$T^{a}_{\underline{n}\underline{l}}$ and $T^{\rho}_{\underline{n}
\underline{l}}$ are the torsions connected with the dreinbein and
gravitino respectively, $\omega^{\underline{m}}$ is the superCartan form.
In the limiting case of the $\theta=0$ our Lagrangian reproduces the
result of ref. \cite{sstv}.

The investigation of the constraints algebra
is analogous to the Maxwell field interaction case. Again due to the
existence of the conventional constraint
$${\cal{R}}_{\alpha\beta}(\hat{X},\Theta)=0$$
the algebra of the second class constraints does not change.

Thus, in the framework of generalized twistor dynamics, superparticle
interacting with background supergravity  "feels" space torsion and, in
the second order over $l$, curvature.

In the end of this section we would like to make some remarks concerning
superparticle interacting with background fields. The demanding
preservation of all original symmetries imposes some conditions on
background fields. In particular, the preservation of the Siegel fermionic
symmetry in the presence of a gauge field requires that the superfield
strength satisfies the super~--~Yang~--~Mills equations of motion (see
ref. \cite{shapiro} and refs.  therein). It is achieved by imposing
conventional constraints (in the case of Maxwell background field these
ones are (15) and (16)).

For the superobject in a general curved superspace, the situation is more
complicated. In the standard formulation of N=1 Brink -- Schwarz
superparticle the action describing the former looks like
$$ S=\int\,d\tau\,{1\over2}e^{-1}\eta_{ab}{\dot{z}}^{M}E_{M}^{a}
{\dot{z}}^{N}E_{N}^{b}$$
and it is easy to note that superparticle does not "feel"  $E^{\alpha}$
components of background supervierbein. For this reason the preserving
original symmetries does not lead to the superfield supergravity equation
of motion but to the classes of equivalence of supergravity background
fields.

In contrast to the standard version twistor -- like superparticle "feels"
all components of background supervierbein and it would be very
interesting to investigate conditions imposed on background supergravity
first steps to the studying of which have been made in refs.
\cite{ds,gal}.

\section{Conclusion.}

In the first time it was assumed to use the twistor shift procedure for the
solution of the bosonic string tachion sector problem. This sector arises
on the particle level when we try to carry out a generalization of twistor
dynamics based on the STVZ action. Unfortunately, there are not any
successes on this direction as well as for attempts of consideration
twistor shift in the case of higher dimensions D=4,6. Thus, the main
application of the twistor shift connected with anyon physics in D=3.

The existence of twistor shift procedure touches upon very important
question concerning the meaning of physical coordinates in general and
connection between worldline (worldsheet) locality and causality with
target space ones.  With consideration of generalized twistor dynamics we
expand phase space by inclusion of additional variables $\lambda$ and
their momenta.  Upon the twistor shift procedure we eliminate additional
phase space coordinates, but it does not mean that we return explicitly to
the original phase space. The absence of clear understanding of this
situation does not allow us to make hard conclusion about, for example,
structure of constraints algebra, which closely connected with
conventional constraints imposed on background fields. We naively assume
that this constraints have an anologous form as in the standard case, but
it may be not really so !

{}From the other point of view the expanded phase
space of a generalized twistor dynamics possesses a noncomutativity of the
space -- time coordinates, restored by twistor shift. However in the last
time there are many consideration of the theories with noncommutative
coordinates on the Plank scale. Of course, the theories of such kind have
many difficulties connected, for example, with the ordering of the
coordinates, introduction of interaction with background fields etc., but
it would be very interesting to investigate a generalized twistor dynamics
without twistor shift in this context.

Thus, a generalized twistor dynamics puts principal and deep questions and
stimulates the finding of their solution.

\vspace{1cm}
{\Large{\bf Acknolegments.}}

The authors are grateful to D.V.Volkov (A.N. personally thank D.V. for the
fruitfull discussions of these problems and great help in preparation of
this article to publication) and also D.P.Sorokin and I.A.Bandos for
stimulating discussions. A.N.  acknowledge
support from International Soros Foundation, Grant N RY 9000 and partial
support from Ukranian State Committee in Science and Technology, Grant N
2.3/664. V.T. thanks to METU (Ankara) for the kind hospitality and T.Aliev
for useful discussions. V.T. acknowledge support from International Soros
Foundation, Grant N U9G000.

\vspace{2cm}

{\Large{\bf Appendix.}}

We choose a real (Majorana) representation for the Dirac matrices
${(\gamma_{m})}_{\alpha}^{\beta}$ and charge conjugation matrix
$C_{\alpha\beta}$ ( D=3 ):
$$\gamma^{0}=C=-i\sigma_{2},\gamma^{1}=\sigma_{1},\gamma^{2}=\sigma_{3},
C_{\alpha\beta}=\varepsilon_{\alpha\beta}$$
where $\sigma_{i}$ are the Pauli matrices. $\gamma$ -- matrices satisfy
the following relations:
$$\{\gamma_{m},\gamma_{n}\}=2g_{mn},\bigskip g_{mn}=diag(-,+,+)$$
$$\sigma_{mn}={1\over4}[\gamma_{m},\gamma_{n}]=-{1\over2}\varepsilon_{mnl}
\gamma^{l}$$
$$2\delta_{\alpha}^{\beta}\delta_{\gamma}^{\delta}=\delta_{\alpha}^{\delta}
\delta_{\gamma}^{\beta}+{(\gamma^{m})}_{\alpha}^{\delta}{(\gamma_{m})}_
{\gamma}^{\beta}$$
where  $\varepsilon^{mnl}$ is the Levi -- Chivita tensor and $(\varepsilon^
{012}=1)$. The vector and spin -- tensor representations connect as
$$x_{\alpha}^{\beta}={(\gamma^{m}x_{m})}_{\alpha}^{\beta}$$
Raising and lowering of spinor indices is carried out by the matrix
$\varepsilon_{\alpha\beta}$ according to the rules:
$$\lambda_{\alpha}=\varepsilon_{\alpha\beta}\lambda^{\beta},
\lambda^{\alpha}=\varepsilon^{\alpha\beta}\lambda_{\beta},
\varepsilon_{12}=-\varepsilon^{12}=-1.$$

\end{document}